\documentclass{article}
\usepackage{spconf,amsmath,graphicx,hyperref,amssymb}
\usepackage{cite}
\usepackage{amsmath,amssymb,amsfonts}
\usepackage{mathrsfs}

\usepackage{graphicx}
\usepackage{subfig}
\usepackage{textcomp}
\usepackage{xcolor}
\usepackage{tikz}
\usetikzlibrary{positioning,arrows.meta}
\usepackage[export]{adjustbox}
\usepackage{algpseudocode}
\usepackage{float}
\usepackage[ruled,vlined]{algorithm2e}

\algrenewcommand\algorithmicrequire{\textbf{Input:}}
\algrenewcommand\algorithmicensure{\textbf{Output:}}
\algnewcommand{\Train}{\item[\textbf{Student training with KD:}]}
\algnewcommand{\IKD}{\item[\textbf{IKD:}]}
\algnewcommand{\RKD}{\item[\textbf{RKD:}]}

\title{Knowledge Distillation for mmWave Beam Prediction \\ Using Sub-6~GHz Channels}
\name{Sina Tavakolian$^{\star}$ \qquad Nhan Thanh Nguyen$^{\star}$ \qquad Ahmed Alkhateeb$^{\dagger}$ \qquad Markku Juntti$^{\star}$\vspace{-0.25cm}}
  
\address{$^{\star}$ Centre for Wireless Communications, University of Oulu, P.O.Box 4500, FI-90014, Finland \\
  $^{\dagger}$School of Electrical, Computer, and Energy Engineering, Arizona State University, AZ, USA\vspace{-0.5cm}}

\begin{document}
%
\setlength{\abovedisplayskip}{3.5pt}
\setlength{\belowdisplayskip}{3.5pt}
    
\maketitle
\begin{abstract}
Beamforming in millimeter-wave (mmWave) high-mobility environments typically incurs substantial training overhead. While prior studies suggest that sub-6~GHz channels can be exploited to predict optimal mmWave beams, existing methods depend on large deep learning (DL) models with prohibitive computational and memory requirements. In this paper, we propose a computationally efficient framework for sub-6~GHz channel-mmWave beam mapping based on the knowledge distillation (KD) technique. We develop two compact student DL architectures based on individual and relational distillation strategies, which retain only a few hidden layers yet closely mimic the performance of large teacher DL models. Extensive simulations demonstrate that the proposed student models achieve the teacher's beam prediction accuracy and spectral efficiency while reducing trainable parameters and computational complexity by 99\%.
\end{abstract}
\begin{keywords}
Knowledge distillation, beam prediction, deep learning, mmWave communications.\vspace{-0.25cm}
\end{keywords}
\section{Introduction}
\label{sec:intro}
\vspace{-0.25cm} 
Future millimeter-wave (mmWave) systems are expected to operate across multiple frequency ranges, including sub-6~GHz and mmWave bands~\cite{8198818}. The spatial correlation between channels in these bands enables the prediction of mmWave beams directly from sub-6~GHz channels~\cite{9121328}. While such mappings are theoretically feasible, deriving them analytically is highly challenging. To address this, deep learning (DL) has been employed to learn these complex relationships~\cite{9121328,10292615}. Specifically, Alrabeiah \textit{et al.}~\cite{9121328} employed a fully-connected deep neural network (DNN) with five hidden layers of 2048 neurons each, achieving high beam prediction accuracy. Vuckovic \textit{et al.}~\cite{10292615} leveraged a convolutional neural network (CNN) with three convolutional layers followed by a fully connected classifier for more efficient representations. However, these large DL models have high computational and memory demands, limiting their practicality.

Recently, knowledge distillation (KD) has emerged as an effective method to enhance the resource efficiency of DL models. KD involves training a large, complex teacher model and then using it to guide the training of a smaller, more efficient student model~\cite{Gou_Yu_Maybank_Tao_2021, MOSLEMI2024100605}. Recent studies have demonstrated the potential of KD in various wireless applications, including transceiver design~\cite{Gao21}, channel estimation and feedback~\cite{Catak22, 9625047, 9789120}, semantic communications~\cite{SemanticKD2023}, user positioning~\cite{10422998}, and remote sensing~\cite{zhang2021learning}.
For mmWave beam prediction, Park \textit{et al.}~\cite{Park11143350, park2025resource} proposed a KD framework, where the teacher model leverages multi-modal sensing data (LiDAR, radar, GPS, and RGB images) as input to train a student model that operates solely on radar data.

In this paper, we explore KD's potential in developing resource-efficient DL models for mmWave beam prediction. We formulate the beam prediction task as a classification problem, where each candidate beam in the predefined mmWave codebook represents a class. The objective is to map sub-6~GHz channel observations to the beam index that maximizes downlink spectral efficiency (SE), thereby avoiding beam training overheads while ensuring that the underlying DL models remain computationally efficient. To this end, we first pretrain a high-capacity teacher model, and then distill its knowledge into compact student models that can be deployed at the base station (BS). We investigate three KD strategies, including individual KD (IKD)~\cite{hinton2015distilling}, relational KD (RKD)~\cite{8954416}, and self-distillation~\cite{selfdistillation}, leading to lightweight DL models for mmWave beamforming. Extensive evaluation shows that the proposed student models achieve nearly 99\% reduction in trainable parameters and floating point operations (FLOPs) compared to the teacher, while maintaining beam prediction accuracy and SE close to that of the teacher.

\vspace{-0.35cm}
\section{System Model and Problem Formulation}
\label{sec:sysmodel}
\vspace{-0.25cm}
\subsection{System Model} 
\vspace{-0.15cm}
We consider a communications system that operates concurrently in the sub-6~GHz and mmWave frequency bands. The system consists of a BS and a user equipment (UE). The BS is equipped with two types of transceivers: one operating in the sub-6~GHz band with \(N_{\text{sub-6}}\) antennas, and the other in the mmWave band with \(N_{\text{mmW}}\) antennas.
At the UE, one antenna is used for sub-6~GHz communications and another for mmWave communications.
A geometric channel model is considered for both sub-6~GHz and mmWave bands~\cite{9121328}.
Let $\mathbf{h}_{\text{sub-6}}[\bar{k}] \in \mathbb{C}^{N_{\text{sub-6}} \times 1}$ denote the sub-6~GHz uplink channel vector at subcarrier $\bar{k}  = 1,2,\ldots,\bar{K}$. 
Similarly, let $\mathbf{h}_{\text{mmW}}[k] \in \mathbb{C}^{N_{\text{mmW}} \times 1}$ denote the mmWave downlink channel vector at subcarrier $k  = 1,2,\ldots,K$.
We assume that the sub-6~GHz antenna array is equipped with a fully digital architecture, which allows obtaining the sub-6~GHz channels based on conventional baseband channel estimation~\cite{7400949}.

For the downlink transmission, the mmWave array at the BS uses analog beamforming to minimize the hardware and power costs. The received signal at the UE on subcarrier $k$ is
\begin{equation}
y_{\text{mmW}}[k] = \mathbf{h}_{\text{mmW}}^{T}[k] \, \mathbf{w} \, s_d + v_{\text{mmW}}[k],
\end{equation}
where $\mathbf{w}$ is the frequency-flat analog beamforming vector selected from a predefined codebook $\mathcal{W}$, $s_d$ is the transmitted data symbol, and $v_{\text{mmW}}[k] \sim \mathcal{N}_{\mathbb{C}}(0,\sigma^2_{\text{mmW}})$ denotes the additive complex Gaussian noise at the receiver. The codebook $\mathcal{W}$ contains $N_{\text{mmW}}$ possible beamforming vectors, enabling the BS to select the optimal beam to maximize the received signal strength at the UE. The downlink achievable rate is
\begin{equation}
R \big(  \mathbf{w} \big) = \sum_{k=1}^{K} \log_2 \left( 1 + \text{SNR} \, \big| \mathbf{h}_{\text{mmW}}^{T}[k] \mathbf{w} \big|^2 \right),
\end{equation}
where the per-subcarrier signal-to-noise ratio (SNR) is defined as $\text{SNR} = P_{\text{mmW}}/(K \sigma^2_{\text{mmW}})$, with $P_{\text{mmW}}$ denoting the total downlink transmit power from the BS.
\vspace{-0.25cm}
\subsection{Problem Formulation} \vspace{-0.15cm}
The optimal beamforming vector \( \mathbf{w}^\star \) that maximizes the downlink achievable rate is determined by
\begin{equation}
\mathbf{w}^\star = \arg \max_{\mathbf{w} \in \mathcal{W}} R \left(  \mathbf{w} \right).
\end{equation}
A direct solution via exhaustive search requires evaluating all beamforming vectors in the codebook $\mathcal{W}$, which is computationally prohibitive for large \(N_{\text{mmW}}\). Moreover, it is worth noting that computing 
$R\!\left(  \mathbf{w} \right)$ 
requires estimating the mmWave channels 
$\{ \mathbf{h}_{\text{mmW}}[k] \}$, which is highly challenging due to the analog architecture of the mmWave array and the lack of uplink pilot signals for mmWave channel estimation.

Alternatively, the problem can be formulated as a beam prediction task~\cite{9121328}, where the goal is to learn a mapping
\begin{equation}\label{eq_approx}
    \Phi: \mathbf{h}_{\text{sub-6}} \mapsto \mathbf{w}^\star,
\end{equation}
that predicts the optimal mmWave beam index directly from the sub-6~GHz channel. This formulation naturally leads to a multi-class classification task, with each beamforming vector in the codebook $\mathcal{W}$ representing a distinct class. The existence of such a mapping relies on two assumptions: (i) the mapping from UE locations to sub-6~GHz channels is bijective, ensuring each channel realization is unique, and (ii) the mmWave codebook provides a unique optimal beam for every channel. Under these conditions, a sufficiently large DNN can approximate the mapping~\eqref{eq_approx} with arbitrarily small error~\cite{9121328}.  

While this demonstrates the potential of DL-based beam prediction, existing models are often computationally demanding, incurring high memory usage and significant inference latency~\cite{9121328}. To address these limitations, in the next section, we develop lightweight DL models based on KD to perform the mapping efficiently.

\vspace{-0.35cm}
\section{KD-Based mmWave Beamforming}
\label{sec:kd}
\vspace{-0.25cm}

The main idea of KD is to transfer knowledge from a complex, high-performing teacher model to a compact student model~\cite{Gou_Yu_Maybank_Tao_2021}, which is then employed for online inference.
To elaborate on how KD is applied for the mapping~\eqref{eq_approx}, we next present the basic formulation and structure of the teacher model, followed by the lightweight student models developed by distilling the knowledge from the teacher.

\vspace{-3.5mm}
\subsection{Training Teacher Model}
\vspace{-2mm}
As the learning task is to approximate the mapping~\eqref{eq_approx}, the teacher model takes the processed sub-6~GHz channel as the input, which is obtained by concatenating the subcarrier-domain channel vectors across all antennas. Specifically, with $N_{\text{sub-6}}$ antennas and $\bar{K}_{\text{sub-6}}$ subcarriers, the input dimension is $d_{\text{in}} = 2N_{\text{sub-6}}{\bar{K}_{\text{sub-6}}}$, where the factor $2$ accounts for representing real and imaginary parts separately. We denote this concatenated input vector by $\mathbf{x} \in \mathbb{R}^{d_{\text{in}}}$, and let $\mathcal{X}$ be the set of all such training samples obtained from $\{\mathbf{h}_{\text{sub-6}}[\bar{k}]\}_{\bar{k}=1}^{\bar{K}_{\text{sub-6}}}$ across different UE locations.
The teacher is trained in a supervised manner.
Specifically, let $\{(\mathbf{x}_i,\mathbf{z}_i)\}_{i=1}^M$ denote a minibatch of $M$ training samples, where $i$ indexes the pair consisting of input $\mathbf{x}_i \in \mathcal{X}$ and its corresponding 
one-hot label $\mathbf{z}_i \in \{0,1\}^{|\mathcal{W}|}$.
For each $\mathbf{x}_i$, the teacher produces logits 
$\ell_{\mathsf{T}}(\mathbf{x}_i) \in \mathbb{R}^{|\mathcal{W}|}$ and probability distribution $\mathbf{p}_i^{\mathsf{T}} = \mathtt{softmax}(\ell_{\mathsf{T}}(\mathbf{x}_i))$ over all the beams in the codebook.
The teacher is trained with the classification loss
$\mathscr{L}_{\text{CE}} = \frac{1}{M}\sum_{i=1}^M \mathtt{CE}(\mathbf{p}_i^{\mathsf{T}}, \mathbf{z}_i)$, where $\mathtt{CE}(\cdot,\cdot)$ denotes the cross-entropy function.
To provide effective guidance for the student models, the teacher must employ a sufficiently large network, as will be specified in Section~\ref{sec:results}.


\vspace{-3.5mm}
\subsection{Training Student Models Based on KD}
\vspace{-2mm}
To achieve low-complexity yet reliable beam prediction, we train lightweight DL models by distilling knowledge from the high-performing teacher, as detailed next.

\textbf{Individual Knowledge Distillation:}
Let $\ell_{\mathsf{S}}(\mathbf{x}_i) \in \mathbb{R}^{|\mathcal{W}|}$ and $
\mathbf{p}_i^{\mathsf{S}} = \mathtt{softmax}\big(\ell_{\mathsf{S}}(\mathbf{x}_i)\big)
$ be the student's logits and the corresponding probability distribution associated with minibatch $\{(\mathbf{x}_i,\mathbf{z}_i)\}_{i=1}^M$.
The teacher and student further produce soft targets, i.e., softened distributions using a temperature parameter $\tau>1$, expressed as
\begin{equation}
\label{softend}
\tilde{\mathbf{p}}_i^{\mathsf{T}} = \text{softmax}\!\left(\tfrac{\ell_{\mathsf{T}}(\mathbf{x}_i)}{\tau}\right), 
\tilde{\mathbf{p}}_i^{\mathsf{S}} = \text{softmax}\left(\tfrac{\ell_{\mathsf{S}}(\mathbf{x}_i)}{\tau} \right).
\end{equation}
The total training loss is obtained by
\begin{equation}
\label{IKD}
\mathscr{L}_{\text{IKD}}\! =\! \frac{1}{M} \!\sum_{i=1}^M 
\!\Big( \alpha \tau^2 \, \underbrace{\mathtt{CE}(\tilde{\mathbf{p}}_i^{\mathsf{S}}, \tilde{\mathbf{p}}_i^{\mathsf{T}})}_{\text{IKD loss}}
+ (1\!-\!\alpha) \, \!\!\underbrace{\mathtt{CE}(\mathbf{p}_i^{\mathsf{S}}, \mathbf{z}_i)}_{\text{classification loss}} \!\Big).
\end{equation}
where $0 \leq \alpha \leq 1$ balances the contribution of distillation and supervised learning. The temperature $\tau$ controls how much information about non-optimal beams is revealed to the student. A higher $\tau$ produces a smoother distribution that highlights similarities between beams. This formulation allows the student to learn not only the correct beamforming vector but also the relative confidence levels assigned to all candidate beams in the codebook $\mathcal{W}$.
\begin{algorithm}[t]
\small
\caption{KD-Based Student Model Training}
\label{alg:kd2e}
\LinesNumbered
\KwIn{Processed sub-6~GHz channel $\{\mathbf{x}_i\}$; one-hot labels $\{\mathbf{z}_i\}$ for mmWave beams; pretrained teacher model;  temperature $\tau$ and weight $\alpha$ for IKD; learning rate $\eta$ and optimizer $\mathtt{Opt}$.}
\KwOut{Optimized student model parameters $\boldsymbol{\Theta}$.}

Initialize student DNN parameters $\boldsymbol{\Theta}$\;
\For{each minibatch $\{(\mathbf{x}_i,\mathbf{z}_i)\}_{i=1}^M$}{
    \eIf{use IKD}{ 
        \tcp{IKD-based training}
        Execute the teacher and student models to obtain logits 
        $\ell_{\mathsf{T}}(\mathbf{x}_i)$ and $\ell_{\mathsf{S}}(\mathbf{x}_i)$, respectively\;
        Compute soft targets via \eqref{softend}\;
        Compute loss $\mathscr{L}_{\text{IKD}}$ via \eqref{IKD}\;
    }{
        \tcp{RKD-based training}
        Extract features $f_{\mathsf{T}}(\mathbf{x}_i)$ and $f_{\mathsf{S}}(\mathbf{x}_i)$ 
        from the teacher and student models for all $\{\mathbf{x}_i\}_{i=1}^M$\;
        
        \For{each pair $(\mathbf{x}_i,\mathbf{x}_j)\in\mathcal{X}^2$}{
            Compute $\psi_{\mathsf{T}}^{\text{dist}}$ and $\psi_{\mathsf{S}}^{\text{dist}}$ via \eqref{eq:psi_dist}\, \;
            Compute Huber loss $g(\psi_{\mathsf{T}}^{\text{dist}}, \psi_{\mathsf{S}}^{\text{dist}})$ via ~\eqref{huber}\;
            Update $\mathscr{L}_{\text{RKD}}^{\text{dist}} \gets \mathscr{L}_{\text{RKD}}^{\text{dist}} 
                    + g(\psi_{\mathsf{T}}^{\text{dist}}, \psi_{\mathsf{S}}^{\text{dist}})$\;
        }

        \For{each triplet $(\mathbf{x}_i,\mathbf{x}_j,\mathbf{x}_k)\in\mathcal{X}^3$}{
            Compute $\psi_{\mathsf{T}}^{\text{angl}}$ and $\psi_{\mathsf{S}}^{\text{angl}}$ via \eqref{eq:psi_angl}\;
            Compute Huber loss $g(\psi_{\mathsf{T}}^{\text{ang}}, \psi_{\mathsf{S}}^{\text{ang}})$ via ~\eqref{huber}\;
            Update $\mathscr{L}_{\text{RKD}}^{\text{angl}} \gets \mathscr{L}_{\text{RKD}}^{\text{angl}} 
                    + g(\psi_{\mathsf{T}}^{\text{angl}}, \psi_{\mathsf{S}}^{\text{angl}})$\;
        }
        Compute loss $\mathscr{L}_{\text{RKD}}$ based on \eqref{eq:total_loss_RKD}\;
    }
    \tcp{Backpropagation and update}
    Compute gradients $\nabla_{\boldsymbol{\Theta}}\mathscr{L}$\;
    Update $\boldsymbol{\Theta} \gets \boldsymbol{\Theta} - \eta \,\mathtt{Opt}\!\big(\nabla_{\boldsymbol{\Theta}}\mathscr{L}\big)$\;
}
\end{algorithm}


\textbf{Relational Knowledge Distillation:}
RKD~\cite{8954416} extends IKD by focusing on the relational structure across sub-6~GHz channel realizations of a UE at different locations.
The similarity between such channel realizations reflects how spatial positions and propagation conditions map into distinct mmWave beams.
To capture the relations, we form all possible $n$-tuples $(\mathbf{x}_1,\ldots,\mathbf{x}_n)$ within each minibatch $\{(\mathbf{x}_i,\mathbf{z}_i)\}_{i=1}^M$ and quantify the relations among the samples inside each tuple. Specifically, the entire minibatch is first passed through the teacher and student networks to obtain features $f_\mathsf{T}^{(r)}(\mathbf{x}_i)$ and $f_\mathsf{S}^{(r)}(\mathbf{x}_i)$, where $r$ indexes the networks' layer. Unlike IKD, which relies solely on final logits, RKD can exploit features from an intermediate layer $r$.

Given the extracted features corresponding to the samples in each tuple, relational structures are then quantified through distance-wise and angle-wise potential functions: the former captures pairwise ($n=2$) similarity between processed sub-6~GHz channels via the normalized Euclidean distance between features, while the latter encodes triplet relations ($n=3$) through the cosine of the angle between feature difference ~\cite{8954416}. These can be mathematically formulated as:
\begin{align}
\psi_\mathsf{A}^{\mathsf{dist}} \,&=\, \tfrac{1}{\mu}\,\| f_\mathsf{A}^{(r)}(\mathbf{x}_i) - f_\mathsf{A}^{(r)}(\mathbf{x}_j) \|_2 , \label{eq:psi_dist}\\
\psi_\mathsf{A}^{\mathsf{angl}} \!&= \cos\!\big\langle f_\mathsf{A}^{(r)}(\mathbf{x}_i)\!-\!f_\mathsf{A}^{(r)}(\mathbf{x}_j),\, f_\mathsf{A}^{(r)}(\mathbf{x}_k)\!-\!f_\mathsf{A}^{(r)}(\mathbf{x}_j) \big\rangle . \label{eq:psi_angl}
\end{align}
where $\mu$ is set to the average pairwise distance over all possible pairs in the minibatch and $\mathsf{A}\in\{\mathsf{T},\mathsf{S}\}$ indicates whether the potentials are computed from the teacher or student features.
The RKD objective is to minimize both the distance-wise and angle-wise losses, expressed as
\begin{align}
\mathscr{L}_{\text{RKD}}^{\text{dist}} 
\!&=\! \!\sum_{\mathcal{X}^2} g\!\left( \psi_{\mathsf{T}}^{\mathsf{dist}}\!, \psi_{\mathsf{S}}^{\mathsf{dist}} \right), 
\mathscr{L}_{\text{RKD}}^{\text{angl}} 
\!=\! \!\sum_{\mathcal{X}^3} g\!\left( \psi_{\mathsf{T}}^{\mathsf{angl}},\! \psi_{\mathsf{S}}^{\mathsf{angl}} \right)\!.
\end{align}
where $\mathcal{X}^2$ is the set of all possible pairs in the minibatch and $\mathcal{X}^3$ the set 
of all possible triplets, and $g(\cdot,\cdot)$ measures the discrepancy between teacher and student relational structures. In this work, we formulate $g(\cdot,\cdot)$ using the Huber loss, i.e.,
\begin{equation}
\label{huber}
g(\psi_{\mathsf{T}},
\psi_{\mathsf{S}}) =
\begin{cases} 
\frac{1}{2} (\psi_{\mathsf{T}} - \psi_{\mathsf{S}})^2 & \text{for } |\psi_{\mathsf{T}} - \psi_{\mathsf{S}}| \leq 1, \\
|\psi_{\mathsf{T}} - \psi_{\mathsf{S}}| - \frac{1}{2} & \text{otherwise.}
\end{cases}
\end{equation}
The total training loss is given by
\begin{equation}
\label{eq:total_loss_RKD}
\mathscr{L}_{\text{RKD}}\!\!\! \;=\;
\!\!\!\underbrace{\frac{1}{M} \sum_{i=1}^M 
   \mathtt{CE}(\mathbf{p}_i^{\mathsf{S}}, \mathbf{z}_i)}_{\text{classification loss}}
+ \!\!\underbrace{\frac{1}{\binom{M}{2}}\mathscr{L}_{\text{RKD}}^{\text{dist}}}_{\text{distance-wise loss}}
\!\!\!+\! \underbrace{\frac{1}{\binom{M}{3}}\mathscr{L}_{\text{RKD}}^{\text{angl}}}_{\text{angle-wise loss}} .
\end{equation}

\begin{figure}[!t]
    \centering\small
    \includegraphics[width=\linewidth]{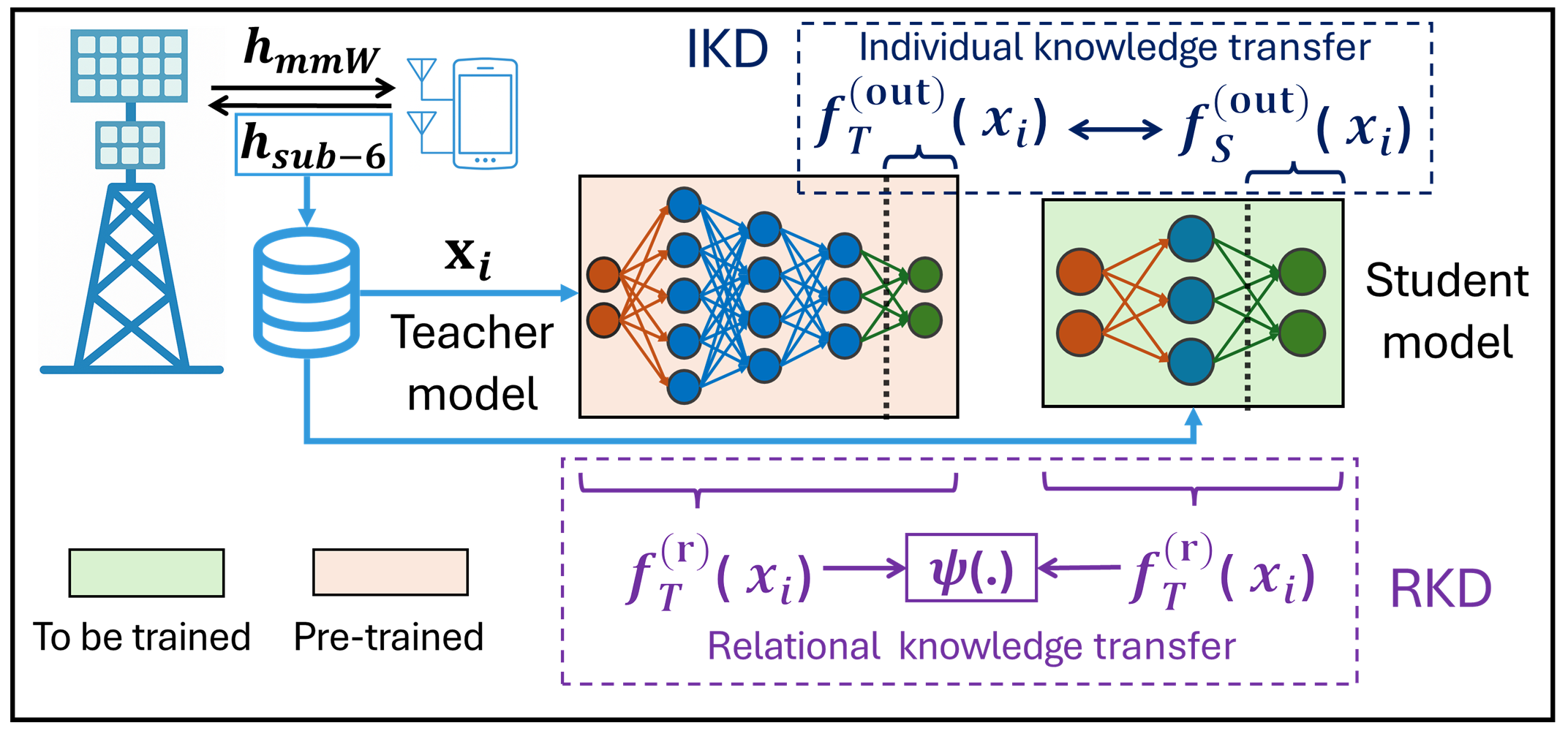}\vspace{-0.25cm}
    \caption{Schematic of the proposed KD framework.\vspace{-0.5cm}}
    \label{fig:KDscheme}
\end{figure}


The KD-based training procedure is summarized in Algorithm~\ref{alg:kd2e}. In IKD, the student learns from both the ground-truth beam labels and the teacher’s softened output distributions, thereby mimicking its predictive behavior. In RKD, the student further preserves the relational structure across sub-6~GHz channels by matching relational potentials with those of the teacher. This process is also illustrated in Fig.~\ref{fig:KDscheme}.

\begin{figure*}[!htb]
    \centering
    \small
    \begin{minipage}{0.32\textwidth}
        \centering
        \includegraphics[width=\linewidth]{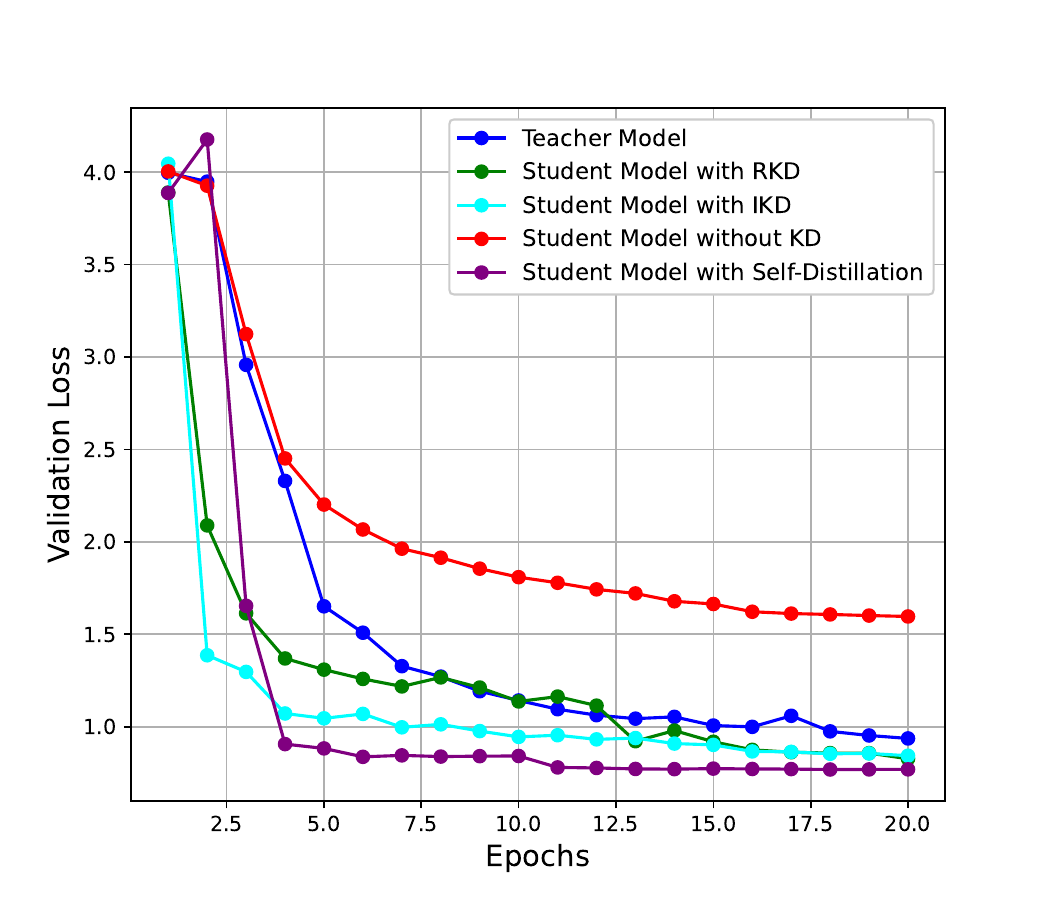}
        \vspace{-8mm}
        \caption{\small Validation loss of the considered models over training epochs.}
        \label{fig3:convergence}
    \end{minipage}\hfill
    \begin{minipage}{0.32\textwidth}
        \centering
        \includegraphics[width=\linewidth]{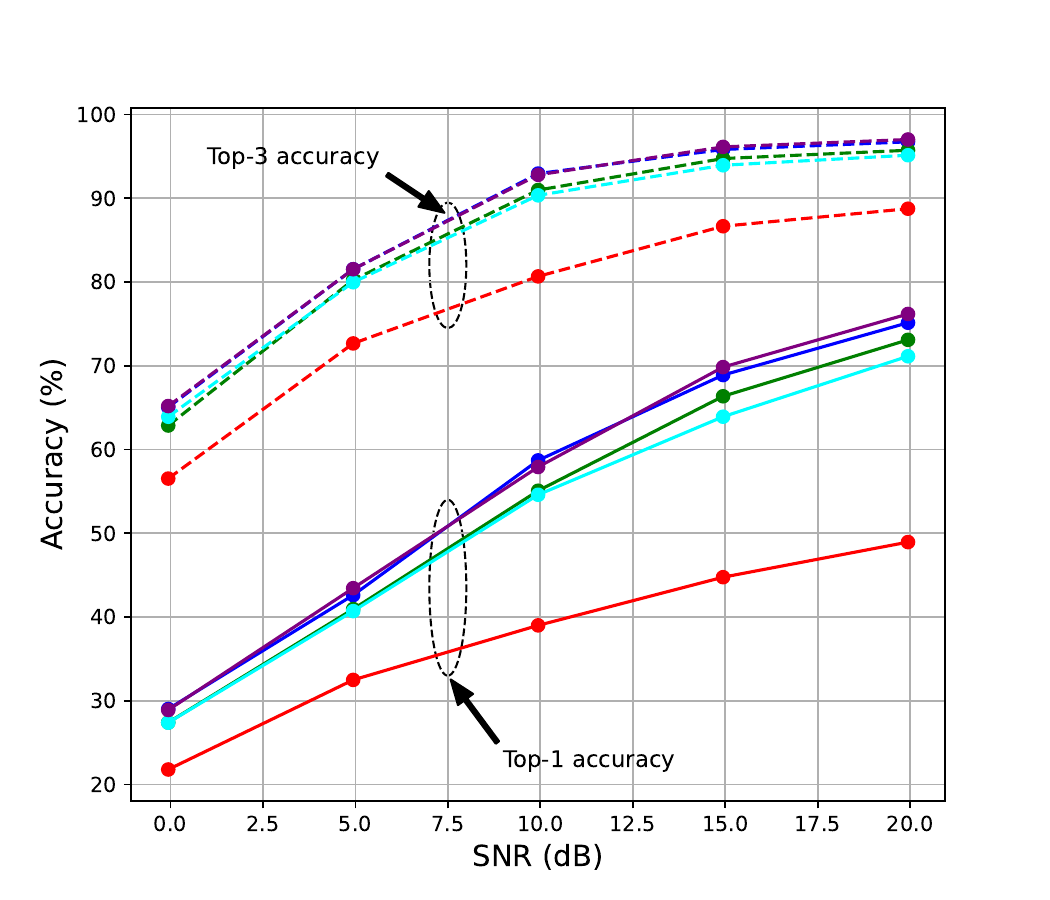}
        \vspace{-8mm}
        \caption{\small Beam prediction accuracies of the compared DL models versus SNRs.}
        \label{fig1:accuracy}
    \end{minipage}\hfill
    \begin{minipage}{0.32\textwidth}
        \centering
        \includegraphics[width=\linewidth]{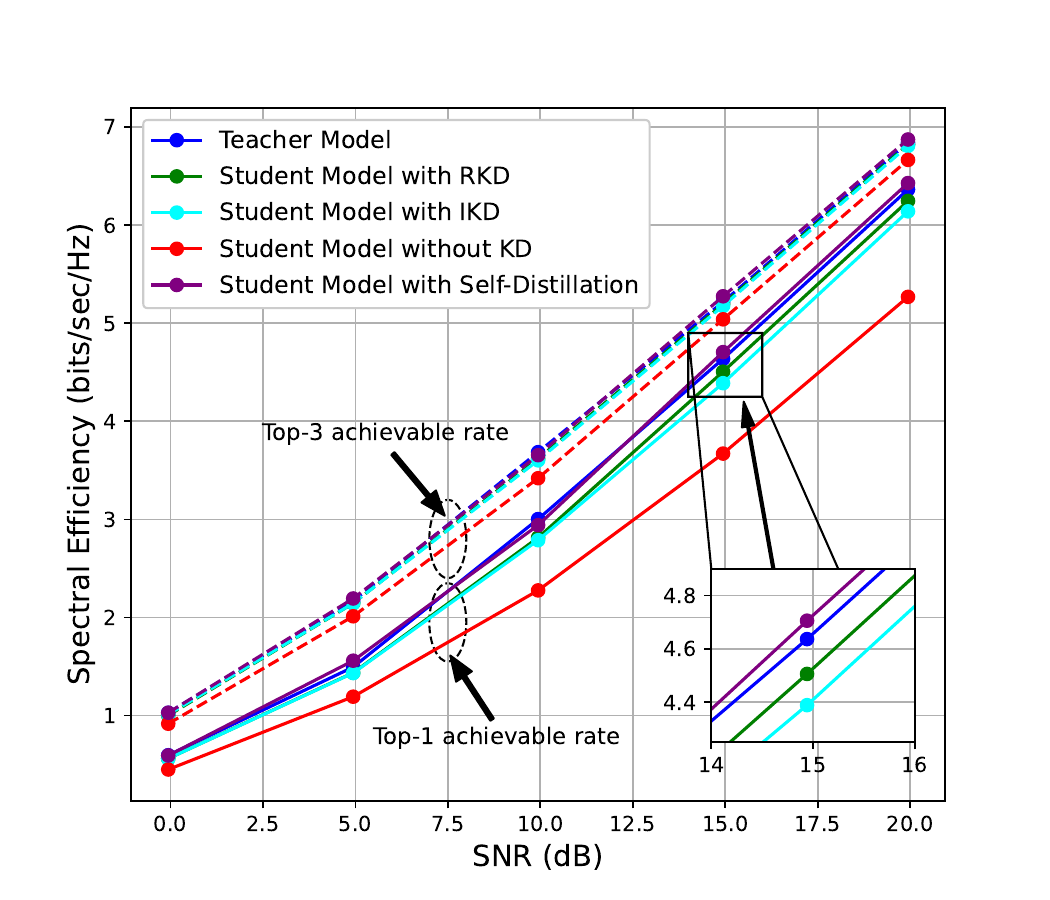}
        \vspace{-8mm}
        \caption{\small SE performance achieved by the considered DL models versus SNRs.}
        \label{fig2:se}
    \end{minipage}
\vspace{-5mm}
\end{figure*}

\vspace{-0.40cm}
\section{Numerical Results}
\vspace{-0.15cm}
\label{sec:results}

\textbf{Dataset and Simulation Settings:} We conduct our simulations using the O1\_28 and O1\_3p5 setups in the O1 scenario in DeepMIMO dataset~\cite{Alkhateeb2019}. 
Both involve a BS (BS 3) serving active UEs. The O1\_28 configuration utilizes 64 antennas with a 0.5~wavelength spacing, a 0.5~GHz bandwidth, and 512 OFDM subcarriers, while O1\_3p5 has 4 antennas, 0.02~GHz, and 32 OFDM subcarriers. 
%
%
For the teacher network, we adopt the fully connected feed-forward DNN architecture with four hidden layers of 1024 neurons each. Notice that, as per~\cite[Table~II]{9121328}, this model achieves performance close to the larger models employed in~\cite{9121328}.
The lightweight student models consist of only two hidden layers with 64 neurons each. 
For IKD, we set $\tau=10$ and $\alpha=0.9$, which provide the best trade-off between the teacher’s softened outputs and the true labels. We also implement a self-distillation approach, where the teacher and student share the same architecture~\cite{selfdistillation}, for comparison. The performance of the considered DL models is evaluated using three metrics: Top-$k$ accuracy of beam prediction, SE of the predicted beam, and the validation loss during training. Here, Top-$k$ accuracy measures whether the ground-truth label lies within the model's Top-$k$ predicted labels.


\textbf{Validation Loss:} We first examine the convergence during training  
the teacher, self-distilled, IKD-, and RKD-based student models, as well as a baseline non-distilled model trained from scratch without KD. As seen in Fig.~\ref{fig3:convergence}, 
the self-distilled student achieves the fastest convergence, reaching a lower validation loss within 12 epochs, whereas the teacher has not converged even after 20 epochs. Both IKD- and RKD-based students also converge faster than the teacher and the non-distilled baseline, attaining lower validation loss. These results show that KD enables student models to benefit from the pretrained teacher, leading to accelerated convergence.

\textbf{Performance of Student Models:} In Fig.~\ref{fig1:accuracy}, we compare the Top-1 and Top-3 beam prediction accuracies of the teacher and student models. 
As expected, the self-distilled student model matches or surpasses the teacher in both Top-1 and Top-3 accuracies.
The IKD- and RKD-based student models outperform the non-distilled one across all SNR levels. Importantly, they retain much of the teacher’s accuracy with far smaller architecture.
For instance, at $\text{SNR}=15$~dB, the teacher achieves a Top-1 accuracy of $68.90\%$, while the RKD and IKD students reach $66.36\%$ and $63.92\%$, respectively, corresponding to $95\%$ of the teacher’s performance. In contrast, the non-distilled student achieves only $44.76\%$ accuracy, i.e., about $20\%$ worse than the distilled students (with the same model size but trained via KD). 

We compare the SE of the considered schemes  
in Fig.~\ref{fig2:se}. The SE is obtained based on the predicted beams; for Top-3 SE, we consider the three highest-scoring beams and select the resulting best SE. 
It is seen that the teacher model and self-distilled student offer the highest SE, with the RKD and IKD students following closely.
For example, at $\text{SNR}=15$~dB, the self-distilled student achieves the highest Top-1 SE of $4.71$~bits/s/Hz, slightly outperforming the teacher. 
The RKD and IKD students follow with $4.51$ and $4.39$~bits/s/Hz, respectively, retaining over $95\%$ of the teacher’s SE. By contrast, the non-distilled student achieves only $3.67$~bits/s/Hz.


\begin{table}[!t]
\centering
\small
\caption{\small Complexity Comparison of the Models}\vspace{-3mm}
\resizebox{\linewidth}{!}{%
\begin{tabular}{|c|c|c|}
\hline
\textbf{Model} & \textbf{Parameters} & \textbf{FLOPs} \\
\hline
Teacher & $P_{\mathsf{T}} = 3,477,568$ & $F_{\mathsf{T}} = 6,940,816$ \\
\hline
Students & $P_{\mathsf{S}} = 24,768$ & $F_{\mathsf{S}} = 49,152$ \\
\hline
\multicolumn{3}{|c|}{\textbf{Reduction (\%)}} \\
\hline
\multicolumn{2}{|c|}{Parameters} & 
$(1 - {P_{\mathsf{S}}}/{P_{\mathsf{T}}})\times 100 \approx 99.29\%$ \\
\hline
\multicolumn{2}{|c|}{FLOPs} & 
$(1 - {F_{\mathsf{S}}}/{F_{\mathsf{T}}})\times 100 \approx 99.29\%$ \\
\hline
\end{tabular}%
}
\vspace{-5.2mm}
\label{tab:complexity}
\end{table}


\textbf{Complexity Reduction:} Table~\ref{tab:complexity} compares the considered schemes by inference complexity and trainable parameters. The inference complexity is dominated by the MLP forward pass and scales linearly with $N_{\text{sub-6}}$, $\bar{K}_{\text{sub-6}}$, and $|\mathcal{W}|$. The student model achieves a complexity reduction of $99.29\%$ relative to the teacher model specified earlier and used in Figs.~\ref{fig3:convergence}--\ref{fig2:se}, as reported in Table~\ref{tab:complexity}. Additional experiments with teacher models containing $2.14\times 10^5$--$1.74\times 10^7$ parameters yield complexity reductions ranging from $88.41\%$ to $99.86\%$.
This highlights KD’s effectiveness in reducing complexity.



\vspace{-0.50cm}
\section{Conclusion}
\vspace{-0.25cm}
We have proposed lightweight DL models for mmWave beam prediction from sub-6~GHz channels leveraging the KD techniques. By distilling a pretrained teacher model into compact student models, we achieve comparable accuracy and SE with up to $99\%$ fewer trainable parameters and significantly lower complexity for inference. Among the two considered KD techniques, RKD performs slightly better than IKD, and they both outperform the non-distilled baseline. The results demonstrate that KD is a powerful tool for achieving computational and memory efficient yet high-performing DL solutions. Future work will investigate KD techniques for sub-6~GHz channel-mmWave beam mapping with dynamic antenna selection and reduced RF chains.


\section{Acknowledgement}

This work was supported by the Research Council of Finland through 6G Flagship Program (grant 369116) and projects DIRECTION (grant 354901), DYNAMICS (grant 24305016), and CHIST-ERA PASSIONATE (grant 359817), by Business Finland, Keysight, MediaTek, Siemens, Ekahau, and Verkotan via project 6GLearn, and in part by the HORIZON-JU-SNS-2023 project INSTINCT (101139161).

\bibliographystyle{IEEEbib}
\bibliography{IEEEabrv,refs}

@STRING{IEEE_J_STSP       = "{IEEE} J. Sel. Topics Signal Process."}

@STRING{IEEE_J_COM        = "{IEEE} Trans. Commun."}

@STRING{IEEE_J_WCOM       = "{IEEE} Trans. Wireless Commun."}

@STRING{IEEE_J_GRS        = "{IEEE} Trans. Geosci. Remote Sens."}

@STRING{IEEE_O_ACC        = "{IEEE} Access"}

@STRING{IEEE_M_COM        = "{IEEE} Commun. Mag."}

@string{ vtc = {Proc. IEEE Veh. Technol. Conf.}}

@string{ spawc = {Proc. IEEE Works. on Sign. Proc. Adv. in Wirel. Comms.}}

@inproceedings{Gao21,
  author    = {Q. Gao and Z. Cao and D. Li},
  title     = {Defensive distillation based end-to-end auto-encoder communication system},
  booktitle = {Proc. IEEE Int. Conf. Computer Commun.},
  pages     = {109--114},
  year      = {2021}
}

@article{park2025resource,
  title={Resource-Efficient Beam Prediction in mmWave Communications with Multimodal Realistic Simulation Framework},
  author={Park, Yu Min and Tun, Yan Kyaw and Saad, Walid and Hong, Choong Seon},
  journal={arXiv preprint arXiv:2504.05187},
  year={2025}
}

@article{Catak22,
  author    = {F. O. Catak and M. Kuzlu and E. Catak and U. Cali and O. Guler},
  title     = {Defensive distillation-based adversarial attack mitigation method for channel estimation using deep learning models in next-generation wireless networks},
  journal   = IEEE_O_ACC,
  volume    = {10},
  pages     = {98191--98203},
  year      = {2022}
}

@article{zhang2021learning,
	title={Learning efficient and accurate detectors with dynamic knowledge distillation in remote sensing imagery},
	author={Zhang, Yidan and Yan, Zhiyuan and Sun, Xian and Diao, Wenhui and Fu, Kun and Wang, Lei},
	journal=IEEE_J_GRS,
	volume={60},
	pages={1--19},
	year={2021},
	publisher={IEEE}
}

@ARTICLE{8198818,
  author={Gonzalez-Prelcic, Nuria and Ali, Anum and Va, Vutha and Heath, Robert W.},
  journal=IEEE_M_COM, 
  title={Millimeter-Wave Communication with Out-of-Band Information}, 
  year={2017},
  volume={55},
  number={12},
  pages={140-146},
  doi={10.1109/MCOM.2017.1700207}}

@ARTICLE{9121328,
  author={Alrabeiah, Muhammad and Alkhateeb, Ahmed},
  journal=IEEE_J_COM, 
  title={Deep Learning for mmWave Beam and Blockage Prediction Using Sub-6 GHz Channels}, 
  year={2020},
  volume={68},
  number={9},
  pages={5504-5518},
  doi={10.1109/TCOMM.2020.3003670}}

@ARTICLE{10292615,
  author={Vuckovic, Katarina and Mashhadi, Mahdi Boloursaz and Hejazi, Farzam and Rahnavard, Nazanin and Alkhateeb, Ahmed},
  journal=IEEE_J_WCOM, 
  title={PARAMOUNT: Toward Generalizable Deep Learning for mmWave Beam Selection Using Sub-6 GHz Channel Measurements}, 
  year={2024},
  volume={23},
  number={5},
  pages={5187-5202},
  doi={10.1109/TWC.2023.3324916}}

@article{Gou_Yu_Maybank_Tao_2021, title={Knowledge distillation: A survey}, volume={129}, number={6}, journal={Int. J. Comput. Vis.}, author={Gou, Jianping and Yu, Baosheng and Maybank, Stephen J. and Tao, Dacheng}, year={2021}, month={Mar}, pages={1789–1819}}

@INPROCEEDINGS{9625047,
  author={Tang, Huaze and Guo, Jiajia and Matthaiou, Michail and Wen, Chao-Kai and Jin, Shi},
  booktitle=vtc, 
  title={Knowledge-distillation-aided Lightweight Neural Network for Massive MIMO CSI Feedback}, 
  year={2021},
  volume={},
  number={},
  pages={1-5},
  doi={10.1109/VTC2021-Fall52928.2021.9625047}}

@ARTICLE{9789120,
  author={Guo, Jiajia and Wen, Chao-Kai and Chen, Muhan and Jin, Shi},
  journal=IEEE_J_COM, 
  title={Environment Knowledge-Aided Massive MIMO Feedback Codebook Enhancement Using Artificial Intelligence}, 
  year={2022},
  volume={70},
  number={7},
  pages={4527-4542},
  doi={10.1109/TCOMM.2022.3180388}}

@ARTICLE{10422998,
  author={Al-Ahmadi, Abdullah},
  journal={IEEE Access}, 
  title={Knowledge Distillation Based Deep Learning Model for User Equipment Positioning in Massive MIMO Systems Using Flying Reconfigurable Intelligent Surfaces}, 
  year={2024},
  volume={12},
  number={},
  pages={20679-20691},
  doi={10.1109/ACCESS.2024.3363088}}

@article{hinton2015distilling,
  title={Distilling the knowledge in a neural network},
  author={Hinton, Geoffrey and Vinyals, Oriol and Dean, Jeff},
  journal={arXiv preprint arXiv:1503.02531},
  year={2015}
}

@INPROCEEDINGS{8954416,
  author={Park, Wonpyo and Kim, Dongju and Lu, Yan and Cho, Minsu},
  booktitle={IEEE/CVF CVPR}, 
  title={Relational Knowledge Distillation}, 
  year={2019},
  volume={},
  number={},
  pages={3962-3971},
  doi={10.1109/CVPR.2019.00409}}

@InProceedings{Alkhateeb2019,
author = {Alkhateeb, A.},
title = {{DeepMIMO}: A Generic Deep Learning Dataset for Millimeter Wave and Massive {MIMO} Applications},
booktitle = {Proc. Inf. Theory
Appli. Workshop (ITA)},
year = {2019},
pages = {1-8},
month = {Feb},
Address = {San Diego, CA}, }

@ARTICLE{7400949,
  author={Heath, Robert W. and González-Prelcic, Nuria and Rangan, Sundeep and Roh, Wonil and Sayeed, Akbar M.},
  journal=IEEE_J_STSP, 
  title={An Overview of Signal Processing Techniques for Millimeter Wave MIMO Systems}, 
  year={2016},
  volume={10},
  number={3},
  pages={436-453},
  doi={10.1109/JSTSP.2016.2523924}}

@inproceedings{selfdistillation,
author = {Pareek, Divyansh and Du, Simon S. and Oh, Sewoong},
title = {Understanding the gains from repeated self-distillation},
year = {2025},
isbn = {9798331314385},
publisher = {Curran Associates Inc.},
address = {Red Hook, NY, USA},
booktitle = {Proc. NeurIPS},
articleno = {249},
numpages = {38},
location = {Vancouver, BC, Canada},
series = {NIPS '24}
}

@article{MOSLEMI2024100605,
title = {A survey on knowledge distillation: Recent advancements},
journal = {Mach. Learn. Appl},
volume = {18},
pages = {100605},
year = {2024},
issn = {2666-8270},
doi = {https://doi.org/10.1016/j.mlwa.2024.100605},
author = {Amir Moslemi and Anna Briskina and Zubeka Dang and Jason Li}
}

@INPROCEEDINGS{Park11143350,
  author={Park, Yu Min and Hassan, Sheikh Salman and Saad, Walid and Hong, Choong Seon},
  booktitle=spawc, 
  title={Cross-Modal Knowledge Distillation for Efficient Radar-Only Beam Prediction in mmWave Communications}, 
  year={2025},
  volume={},
  number={},
  pages={1-5},
  doi={10.1109/SPAWC66079.2025.11143350}}

@ARTICLE{SemanticKD2023,
  author={Liu, Chenguang and Zhou, Yuxin and Chen, Yunfei and Yang, Shuang-Hua},
  journal=IEEE_J_WCOM, 
  title={Knowledge Distillation-Based Semantic Communications for Multiple Users}, 
  year={2024},
  volume={23},
  number={7},
  pages={7000-7012},
  doi={10.1109/TWC.2023.3336941}}

\end{document}